\UseRawInputEncoding
\documentclass[pra,superscriptaddress,twocolumn,aps,amssymb,amsmath]{revtex4-1}
\usepackage[utf8]{inputenc}
\usepackage{tikz}
\usetikzlibrary{quantikz2}
\usepackage{graphicx}% Include figure files
\usepackage{dcolumn,multirow}% Align table columns on decimal point
\usepackage{bm}% bold math
\usepackage{hyperref}% add hypertext capabilities

\hypersetup{
    colorlinks=true,
    linkcolor=blue,
    filecolor=magenta,      
    urlcolor= magenta,
    citecolor = red,
    pdftitle={Expedited Noise Spectroscopy of Qubits},
    pdfpagemode=FullScreen,
    }
    
\begin{document}

\preprint{APS/123-QED}
\title{Expedited Noise Spectroscopy of Transmon Qubits}

\author{Bhavesh Gupta}
\thanks{Equal contribution to this work.}
\affiliation{%
Department of Physics, Indian Institute of Technology Madras, Chennai - 600036, India}
\affiliation{Center for Quantum Information, Communication and Computing, Indian Institute of Technology Madras, Chennai - 600036, India}

\author{Vismay Joshi}
\thanks{Equal contribution to this work.}
\affiliation{%
Department of Physics, Indian Institute of Technology Madras, Chennai - 600036, India}
\affiliation{Center for Quantum Information, Communication and Computing, Indian Institute of Technology Madras, Chennai - 600036, India}

\author{Udit Kandpal}
\affiliation{%
Department of Physics, Indian Institute of Technology Madras, Chennai - 600036, India}
\affiliation{Center for Quantum Information, Communication and Computing, Indian Institute of Technology Madras, Chennai - 600036, India}

\author{{Prabha Mandayam}}
\affiliation{Department of Physics, Indian Institute of Technology Madras, Chennai - 600036, India}
\affiliation{Center for Quantum Information, Communication and Computing, Indian Institute of Technology Madras, Chennai - 600036, India}

\author{Nicolas Gheeraert}
\affiliation{Department of Physics, Indian Institute of Technology Madras, Chennai - 600036, India}
\affiliation{School of Interwoven Arts and Sciences (SIAS), Division of Sciences, Krea University, Sri City - 517646, India}

\author{Siddharth Dhomkar}
\email{sdhomkar@physics.iitm.ac.in}
\affiliation{Department of Physics, Indian Institute of Technology Madras, Chennai - 600036, India}
\affiliation{Center for Quantum Information, Communication and Computing, Indian Institute of Technology Madras, Chennai - 600036, India}

\date{\today}

\begin{abstract}

There has been tremendous progress in the physical implementation of quantum protocols in recent times, bringing us closer than ever to realizing the promise of quantum computing. However, environmental noise continues to pose a crucial challenge to scaling up present-day quantum processors. In the presence of uncontrollable noise sources, decoherence limits the qubits’ ability to store information for long periods. Conventional noise spectroscopy protocols can characterize and model environmental noise but are usually resource-intensive and lengthy. Moreover, the noise can vary in time, making the slow extraction futile as the proﬁle cannot be harnessed to perform error mitigation or correction. Here, this challenge is addressed using a machine learning-based methodology that outputs the noise associated with transmon qubits with minimal absolute error. The procedure involves implementing undemanding dynamical decoupling sequences to record coherence decays of the qubits and then predicting the underlying noise spectra with the help of a convolutional neural network pre-trained on a synthetic dataset. While the protocol is virtually hardware-agnostic, its eﬀectiveness is validated using superconducting qubits available on the IBM Quantum platform. These rapidly obtained, yet accurate, noise spectra are further used to design bespoke dynamic decoupling sequences and perform time-dependent noise spectroscopy

\end{abstract}

\maketitle

\section{\label{sec:level1}INTRODUCTION}

Decoherence \cite{Zurek_2003, Maximilian_2005, Ithier_2005} refers to the loss of coherence in a quantum system due to unavoidable interactions with its environment. It remains the Achilles heel in developing practical quantum technologies such as quantum computation \cite{Putterman_2025, Dalzell_2023}, and quantum sensing \cite{Dolde_2011, RevModPhys.89.035002}. Understanding and controlling noise leading to decoherence in quantum systems is therefore an important task in the context of building robust and scalable quantum processors. To this end, one needs to understand the nature of this noise, identify its possible sources, and eﬃciently curb it with the available tools.

Characterizing and modeling noise using conventional spectroscopic methods, requires abundant resources. The way to extract a spectrum of noise that aﬀects a system is based on deconvolution of the spectral overlap of the noise power spectral density function and the corresponding quantum control protocol used to probe the system \cite{Cywinski}. This control method can also be a dynamical error suppression protocol, such as dynamical decoupling (DD) \cite{Hahn_1950, Carr_1954, MG1958, Lidar_2005, Uhrig_2011, Medford_2012, Bar-Gill_2012, Fabbri_2018}, which is a general method to preserve spin coherence in the presence of noise. The control acts as a ﬁlter, allowing one to probe a speciﬁc region of the broad frequency spectrum aﬀecting the quantum system. This can be modeled analytically with the help of the ﬁlter function formalism \cite{Cywinski, Uys_2009}. Conversely, understanding the spectrum of the noise acting on a qubit crucially underpins the optimization of DD protocols that can mitigate such noise. In reality, the noise spectrum varies substatially across diﬀerent qubits in complex and unpredictable ways that are diﬃcult to accurately extract from standard measurements. Therefore, it is hard to determine in advance, which of the several possible DD protocols would oﬀer optimal suppression of decoherence. One could imagine developing a sequence tailored to a particular qubit, but this is achievable only if detailed and accurate understanding of the underlying qubit noise spectrum can be obtained. Furthermore, the design and implementation of eﬃcient quantum error correction (QEC) \cite{terhal2015} protocols and fault-tolerant quantum circuits \cite{Sascha_24}, can improve signiﬁcantly with an understanding of the speciﬁc noise aﬀecting the qubits \cite{leung1997approximate, tuckett2019tailoring, Chou2024, noise-adapted_nM, dutta2024smallest}; rapid extraction of noise spectra thus becomes an imperative tool for both error mitigation and error correction.

Extracting noise spectra rapidly for qubits presents several challenges. While standard quantum noise spectroscopy (QNS) protocols based on Pauli measurements have enabled signiﬁcant progress in characterizing both single- and multi-qubit noise correlations, \cite{PhysRevLett.116.150503,PhysRevA.95.022121,Sung_2019,Sung2021,PRXQuantum.1.010305,PhysRevLett.107.230501} these methods are generally time-intensive and require complex post-processing. Additionally, conventional QNS protocols are highly eﬀective for stationary noise environments, their extended measurement times can lead to a distorted representation of the actual noise spectrum in the presence of time-dependent noise. Furthermore, other complex pulse sequences with minimal spectral leakage can allow noise extraction with relatively fewer experimental runs; \cite{PhysRevA.98.032315} however, these strategies require unique resources and usually have reduced sensitivity. To mitigate these challenges, our approach leverages standard experimental protocols, such as Carr-Purcell-Meiboom-Gil (CPMG) sequences \cite{Hahn_1950, Carr_1954, MG1958} and utilizes a convolutional neural network to rapidly and robustly infer the underlying noise spectrum from measured coherence decay curves. Owing to their success in diverse domains, there has been a thrust to utilize the power of artiﬁcial neural networks to characterize and correct the deleterious environmental eﬀects. \cite{PRXQuantum.2.010316,PhysRevApplied.18.024004,PhysRevA.105.022605,Canonici_2023,Lu_2025} We note that Youssry et al.\cite{Youssry_2020} have devised a machine learning based ‘graybox’ approach that focuses on the characterization of the dynamics of a two-level open quantum system. However, the process of generating training data and implementing the approach on realistic hardware is non-trivial. In this study, we extend a distinct neural network driven methodology \cite{PRXQuantum.2.010316,PhysRevApplied.18.024004} to rapidly predict the time-varying noise spectral densities associated with the state-of-the-art transmon qubits. The technique allows us to potentially capture transient features and minimize temporal averaging, thereby providing a more accurate reﬂection of the instantaneous noise aﬀecting the qubit. This is speciﬁcally useful for fast and scalable device benchmarking, complementing existing QNS techniques. Moreover, we propose a proof-of-principle method to mitigate the environmental noise by constructing customized dynamical decoupling sequences.

The rest of the paper is organized as follows. In Section \ref{sec:level2}, we discuss the mathematical framework for modeling noise in quantum systems and highlight the importance of accurately discerning the environmental noise. We also recall the link between the extracted noise spectrum and the error rates corresponding to various noise sources \cite{Bylander_2011}. Section \ref{sec:level3} details our methodology. Therein, we demonstrate how a deep neural network can be trained to extract the noise spectrum from typical time dynamics measurements such as Hahn echo and CPMG. We then provide an overview of the experimental data processing and DD optimization procedure. In Section \ref{sec:level4}, we benchmark our model’s eﬀectiveness in extracting and mitigating the noise for the IBM Quantum superconducting quantum architecture. Finally, in Section \ref{sec:level5}, we conclude with critical discussion and propose future directions.

\section{\label{sec:level2}Significance of the `Decoherence Functional'}
\begin{table*}[t]
    \centering
    \begin{tabular}{ |p{2.15cm}|p{5.2cm}|p{5cm}|p{4.25cm}| }
        \hline
        \multicolumn{1}{|c|}{\textbf{Sources}} & \multicolumn{1}{c|}{\textbf{Arises from}} & \multicolumn{1}{c|}{\textbf{Noise mechanisms}} & \multicolumn{1}{c|}{\textbf{S(\(\omega\))}} \\
        \hline
        Charge Noise & Charge fluctuators present in the defects or charge traps that reside in interfacial dielectrics, the junction tunnel barrier, and in the substrate itself. & Dominant: Longitudinal (\(T_1\)), Sometimes low frequency ($T_\phi$) & At low frequency \(\propto 1/\omega\), at high frequency \(\propto \omega\) \\
        \hline
        Magnetic Flux Noise & Stochastic flipping of qubits that reside on the surfaces of the superconductors, resulting in random fluctuations of effective-\(\vec{B}\) that biases flux-tunable qubits. & Longitudinal (\(T_1\)), Transverse (\(T_\phi\)) & \(\propto 1/\omega\) \\
        \hline
        Photon Number Fluctuations & In resonator, residual microwave fields in the cavity have photon number fluctuations. & Transverse (\(T_\phi\)) & \(\propto \frac{k}{\omega^2+k^2}; k:\) resonator decay rate \\
        \hline
        Quasiparticles & Unpaired electrons tunneling through a qubit junction. & Longitudinal (\(T_1\)), Transverse (\(T_\phi\)) & \(\propto \frac{N_{qp}\tau_{qp}}{\omega^2+\tau_{qp}^2}; (N_{qp},\tau_{qp}):\) quasiparticle (population, lifetime) \\
        \hline
        Local Two-Level Systems (TLS) & Electric dipole moments resonantly absorb energy from the oscillating \(\vec{E}\) of the qubit mode, and efficiently dissipate it into the phonon or quasiparticle bath. & Longitudinal (\(T_1\)), Transverse (\(T_\phi\)) & Low freq.: White noise, \(1/\omega\); High freq.: Lorentzian type \\
        \hline
    \end{tabular}
% \end{table*}

    \caption{Summary of common noise sources in transmon systems \cite{Krantz_2019}, their frequency dependencies, and noise mechanisms.}%The noise sources include charge noise, magnetic flux noise, photon number fluctuations, quasiparticles, and local two-level systems (TLS), each contributing to longitudinal relaxation and pure dephasing times, \(T_1\) and \(T_\phi\) respectively, depending on their specific interactions.}
    \label{table_1}
    % \end{center}
    \end{table*}
    
 The Lindblad master equation offers a powerful and elegant way to describe the dynamics of open quantum systems~\cite{breuer2002}. This formalism translates the complex interplay between system and environment into a set of linear, non-Hermitian operators known as Lindblad operators, which capture the essence of various noise processes such as relaxation and dephasing~\cite{manzano2020}. If the noise sources are weakly coupled to the qubits and have short correlation times with respect to the system dynamics, the relaxation processes are characterized by two rates~\cite{pradilla_24},  namely,

\begin{align*}
&\text{Longitudinal relaxation rate:} \, \Gamma_1 = \Gamma_{1\uparrow}+\Gamma_{1\downarrow}\equiv \frac{1}{T_1} \nonumber\\
&\text{Transversal relaxation rate:} \, \Gamma_2 = \frac{\Gamma_1}{2}+\Gamma_\phi \equiv \frac{1}{T_2},
\end{align*}

\noindent  Here, the $T_1$ time corresponds to the energy relaxation process, while $T_\phi$ represents pure dephasing, $\Gamma_{1\uparrow}$ ($\Gamma_{1\downarrow}$) denotes excitation rate (damping rate) and  $\Gamma_\phi = 1/T_\phi$ is the pure dephasing rate. In transmon systems, $\Gamma_{1\uparrow} \rightarrow 0$, as they are operated at millikelvin regimes. The total decoherence time $T_2$ reflects the loss of coherence caused by both energy relaxation and pure dephasing. The noisy density matrix $\rho_{\rm noisy}$ for the qubit after the impact of noise can then be written as~\cite{Nielsen_Chuang}, 

\begin{align}\label{density matrix}
    \rho = \begin{pmatrix}
        |\alpha|^2 & \alpha \beta^* \\ \alpha^*\beta & |\beta|^2  \end{pmatrix}& \xrightarrow{\rm  noise}
        \rho_{\rm  noisy}=\nonumber \\ &\begin{pmatrix}
        1+(|\alpha|^2-1)e^{-\Gamma_1 t} & \alpha \beta^*e^{-\Gamma_2 t}\\
        \alpha^*\beta e^{-\Gamma_2 t} & |\beta|^2e^{-\Gamma_1 t}
    \end{pmatrix}
\end{align}
 
 In this weak coupling limit with short correlation times, the phase decay function is simply given by $e^{-\Gamma_{2}t}$. Furthermore, it is important to note that the $T_1$ relaxation noise process is incoherent  and non-unitary,  thus making it irreversible via purely unitary operations. However,  the dephasing noise corresponding to $T_\phi$ can be mitigated by carefully designing  appropriate control pulse sequences.

Table \ref{table_1} presents  the common noise sources in  a transmon architecture, highlighting the types of coupling between the noise and qubit axes that contribute to decoherence across different frequency regimes. In superconducting qubits, the broadband dephasing noise  -- including, for example, flux noise, charge noise, and critical-current noise --  tends to exhibit a $1/f$-like power spectrum,  as mentioned in Table \ref{table_1}. Such noise is singular near $\omega=0$, has long correlation times, and generally \textit{does not fall} within the Bloch-Redfield description \cite{Krantz_2019}. In this case, the decay function of the off-diagonal terms in Eq.~\eqref{density matrix} is generally non-exponential, and for such cases, the simple expression $\Gamma_2=\Gamma_1/2+\Gamma_\phi$ is \textit{not applicable}.

For $1/f$ noise spectra, under free evolution, the phase decay function is itself a Gaussian $\exp[-(t/T_{\phi,G})^2]$ (stretched exponential decay) \cite{Krantz_2019}. Furthermore, this function is separable from the $T_1$-type exponential decay because the $T_1$-noise remains regular at the qubit frequency. Therefore, the modified density matrix after the action of noise is now given by \cite{Krantz_2019},
\begin{equation}\label{modified}
    \rho=\begin{pmatrix}
        1+(|\alpha|^2-1)e^{-\Gamma_1 t} & \alpha\beta^*e^{-\frac{\Gamma_1}{2}t}e^{-\chi(t)} \\ \alpha^*\beta e^{-\frac{\Gamma_1}{2}t}e^{-\chi(t)} & |\beta|^2e^{-\Gamma_1 t}
    \end{pmatrix}.
\end{equation}
Here, the decay function $e^{-\chi(t)}$ is characterized by the \textit{decoherence functional} $\chi(t)$, which generalizes pure dephasing to include non-exponential decay functions. Because the function is no longer purely exponential, we cannot formally write the transverse relaxation decay function as $e^{-t/T_2}$.

The decoherence functional $\chi(t)$ is related to the noise spectrum or power spectral density (PSD) $S(\omega)$ as follows.
\begin{equation}\label{inteq}
\chi(t)=- \ln C(t) =  \int_{0}^{\infty}\frac{d\omega}{\pi}S(\omega)\frac{F(\omega t)}{\omega^{2}}.
\end{equation}
\noindent Here, $C(t)$ denotes the coherence curve and $F(\omega t)$ is the filter function associated with a given pulse sequence \cite{Cywinski_2008}. In addition to $1/f$-type dephasing mechanisms, there are also ``white" pure dephasing mechanisms, which give rise to an exponential decay function for the dephasing component of $T_2$.  The noise spectrum $S(\omega)$  thus exclusively characterizes the dephasing mechanism.

It is important to note  here that  the noisy density matrix in Eq.~\eqref{density matrix} is equivalent to what one may obtain using the Kraus operator-sum description of noise in quantum systems~\cite{Nielsen_Chuang}. The Kraus operators describe the decoherence resulting from the interaction between the system and its environment over a period of time. The form of the Kraus operators takes into account the decoherence functional, as explained in the Supplementary Information (SI). The density matrix in Eq.~\eqref{density matrix} is thus identical to the density matrix in SI Eq.(3)  obtained due to the combined effect of the dephasing and amplitude-damping channels on the initial density matrix, assuming that the dephasing probability $p(t)$ is simply given by $p=\frac{1}{2}(1+e^{-\Gamma_\phi t})$. 
 
 More generally, the  dephasing probability $p(t)$  maybe expressed as,
\begin{equation}
    p (t) = \frac{1}{2}(1+e^{-\chi(t)}).
\end{equation}
\noindent In this case, the noisy density matrix in SI Eq.(3) gets modified to Eq.~\eqref{modified}. This elucidates the connection between the two different approaches by demonstrating how the decoherence functional can be utilized to derive the explicit probabilities within the Kraus operator framework. 

This connection makes a more precise simulation of noisy dynamics possible, bridging the gap between theoretical descriptions and practical implementations of quantum systems under the influence of noise. Therefore, critical information on qubit dynamics can be derived from the noise spectral density, $S(\omega)$,  which characterizes the environment. 
% Extracting $S(\omega)$ from the measured signal, however, is not straightforward, as the process typically involves a deconvolution prone to error. 
The traditional approaches to noise spectroscopy require deconvolving the measured coherence decay $C(t)$ with the known filter function $F(\omega t)$. In other words, one has to solve an integral equation that is mathematically ill-posed and, thus, does not guarantee an accurate and unique solution. Nonetheless, this is a typical inverse problem where $C(t)$ can be readily obtained from a known $S(\omega)$. Artificial neural networks excel at solving problems such as these \cite{PRXQuantum.2.010316,PhysRevApplied.18.024004,Mohapatra2025}, where they can find a well-approximated function to perform the deconvolution, as explained in the subsequent section. The accuracy and the uniqueness of the solution in this case rely on the constraints placed on the $S(\omega)$, however, a large parameter space can still be incorporated to retain the generality.
%Recent work \cite{PhysRevApplied.18.024004} suggests that this problem can be largely mitigated through the use of deep feedforward neural networks as explained in the subsequent section.

\section{\label{sec:level3}Noise Spectroscopy \& Mitigation Methodology}
The proposed methodology is designed specifically to extract the noise spectrum $S(\omega)$ associated with $T_\phi$ relaxation noise. Experimentally, the $T_\phi$ decay function is isolated by removing the $T_1$ contribution from the overall $T_2$ decay. While $T_1$ relaxation can be addressed primarily through improved qubit fabrication and noise-adapted QEC techniques~\cite{jayashankar2023, ofek2016extending}, our focus here is on estimating and mitigating dephasing noise.

The key element of our approach is a neural network trained to produce the noise spectrum \( S(\omega) \) affecting a given qubit when it is provided with the coherence decay function \( C(t) \) of that qubit as an input. We assume that $S(\omega)$ is stationary, Gaussian, and couples exclusively along the qubit’s $z$-axis, which is a well-studied characteristic of dephasing noise. By leveraging the information from the noise spectrum, we then optimize the DD pulse sequence used to probe the $T_2$ decay. Our method is schematically shown in Fig.~\ref{training_network} and comprises the following steps:

\begin{enumerate}

    \item \textit{Training Data Generation}: Based on the previous work \cite{Krantz_2019}, we assumed that the noise spectra $S(\omega)$ follow some complex yet well-defined functional form. Particularly, we considered that white noise dominates at relatively low frequencies, $1/\omega$-type noise becomes prominent at intermediate frequencies, and $k/(k^2+\omega^2)$-type noise takes over at relatively high frequencies. Thus, we generated tens of thousands of noise spectral densities $S(\omega)$ by varying noise amplitudes and frequency cutoffs associated with the aforementioned noise types. Additionally, the stitched noise spectra were smoothed to avoid unrealistic discontinuities. We then used Eq.~\eqref{inteq} to evaluate the corresponding decoherence curves $C(t)$ by convoluting the generated noise spectra with the CPMG filter function having 32 $\pi$-pulses (CPMG-32). Finally, we add random Gaussian noise to the synthetic coherence curves $C(t)$ to emulate the cumulative impact of experimental imperfections, collectively referred to as \textbf{experimental noise}. More details on data generation can be found in the SI. 
    % \ref{sec:level6.2}.
    
    \item \textit{Network Construction and Training}: 
    Using the numerically generated noise spectral densities and their associated coherence curves $C(t)$, we trained a convolution neural network \cite{PhysRevApplied.18.024004} to identify the noise spectral density $S(\omega)$ based on a single decoherence curve $C(t)$ provided at the network input. A detailed description is provided in the SI.
    % \ref{sec:level6.2}.
    \item \textit{Experimental Data Acquisition}:
    For experimental validation, we performed measurements on IBM’s superconducting quantum processors using the Qiskit package. After determining the optimal amplitude for the square-shaped $\pi$-pulse, we performed sequential $T_1$ and $T_2$ decay experiments using a CPMG-32 pulse protocol. The choice of CPMG-32 sequence over conventional Hahn-echo protocol was made to enable probing of the complex high-frequency region of the noise spectrum, which lies beyond the white noise-dominated low-frequency region. Using the $T_1$ and $T_2$ data we generated the pure dephasing $T_{\phi}$ decay as per Eq.~\eqref{coherence}, where $P_0$ and $P_1$ are the probabilities of $|0\rangle$ (for $T_2$) and $|1\rangle$ (for $T_1$):
    \begin{equation}\label{coherence}
    C(t)= e^{-\chi(t)} \sim \frac{e^{-\Gamma_2 t}}{\sqrt{e^{-\Gamma_1 t}}} = \frac{P_0(T_2)}{\sqrt{P_1(T_1)}}
    \end{equation}
    
    \item \textit{Noise Spectra Prediction}: The noise spectrum prediction is then obtained almost instantaneously, by providing the acquired experimental data as an input to our trained neural network.
    
    \item \textit{Optimization of DD Pulses}:
    The optimization begins by defining the filter function $F(\omega t)$ corresponding to a given DD sequence. By substituting the extracted noise spectrum $S(\omega)$ into Eq.\eqref{inteq}, one can effectively obtain an objective function that seeks to minimize $\chi(t)$. The goal is to find the optimal timing of pulses that reduces the overlap between $S(\omega)$ and $F(\omega t)$. We obtained nearly optimal control pulse sequences by using the SciPy optimizer \cite{2020SciPy-NMeth} -- the Sequential Least SQuares Programming (SLSQP) algorithm \cite{kraft1988software} to optimize the pulse sequence. Subsequently, the customized protocols were implemented on the investigated qubits to validate their fidelity.   
\end{enumerate}

\section{\label{sec:level4}Results and Discussion}

\subsection{\label{sec:level4.1}Network Performance on the Test Data}

\begin{figure*}[ht]
    \centering
    \includegraphics[width=7in]{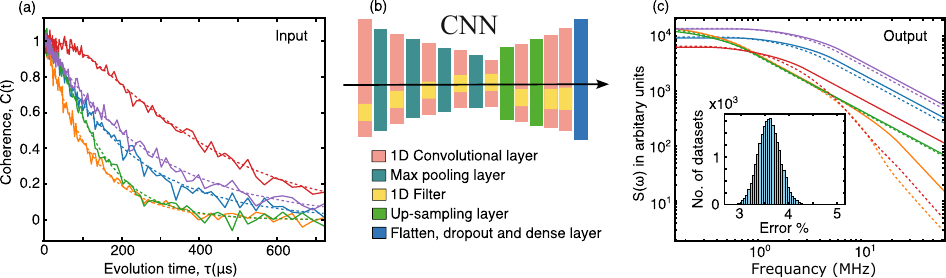}
    \caption{(a) Randomly chosen 5 test input decoherence curves as functions of evolution time; the dashed lines denote the curves evaluated from the predicted $S(\omega)$. (b) A schematic diagram of the neural network developed in this work. (c) The numerically generated (solid lines) and the predicted (dashed line) noise spectra associated with the input decoherence curves. The inset shows a histogram of the estimation errors in coherence curves. }
    \label{training_network}
\end{figure*}

We trained the neural network on a subset of the generated pairs of decoherence curves $C(t)$ and noise spectra $S(\omega)$ (the `training' set). To then test the ability of the trained network to produce the correct noise spectrum from a single decoherence curve, we used another subset of the data (the `test' set) and compared the noise spectrum produced by our network with the originally generated one. The solid lines in Fig.~\ref{training_network}(c) represent the `test' noise spectra, which were used to obtain the corresponding `test' coherence curves, shown in solid lines in Fig.~\ref{training_network}(a), by applying Eq.~\ref{inteq}. The dashed lines in Fig.~\ref{training_network}(c) are then the noise spectra obtained by applying the trained model to the `test' coherence curves. As a final check, the neural net-produced noise spectra were used to obtain decoherence curves to see how well those match the `test' decoherence curves that had been given as an input to the network. These are plotted in dashed lines in Fig.~\ref{training_network}(a), and indeed show good agreement with the `test' decoherence curves, while demonstrating the de-noising capabilities of our network. The accuracy of the model prediction is excellent with a mean absolute error of 3.6\%, as shown in the inset. Note that this error is largely limited by the random noise added to the training data to mimic the experimental signal-to-noise ratio, suggesting that the performance of the network is near optimal

\subsection{\label{sec:level4.2}Error Analysis and Robustness of Neural Network-Based Noise Spectroscopy}
\begin{figure}[!ht]
    \centering
    \includegraphics[width=0.8\linewidth]{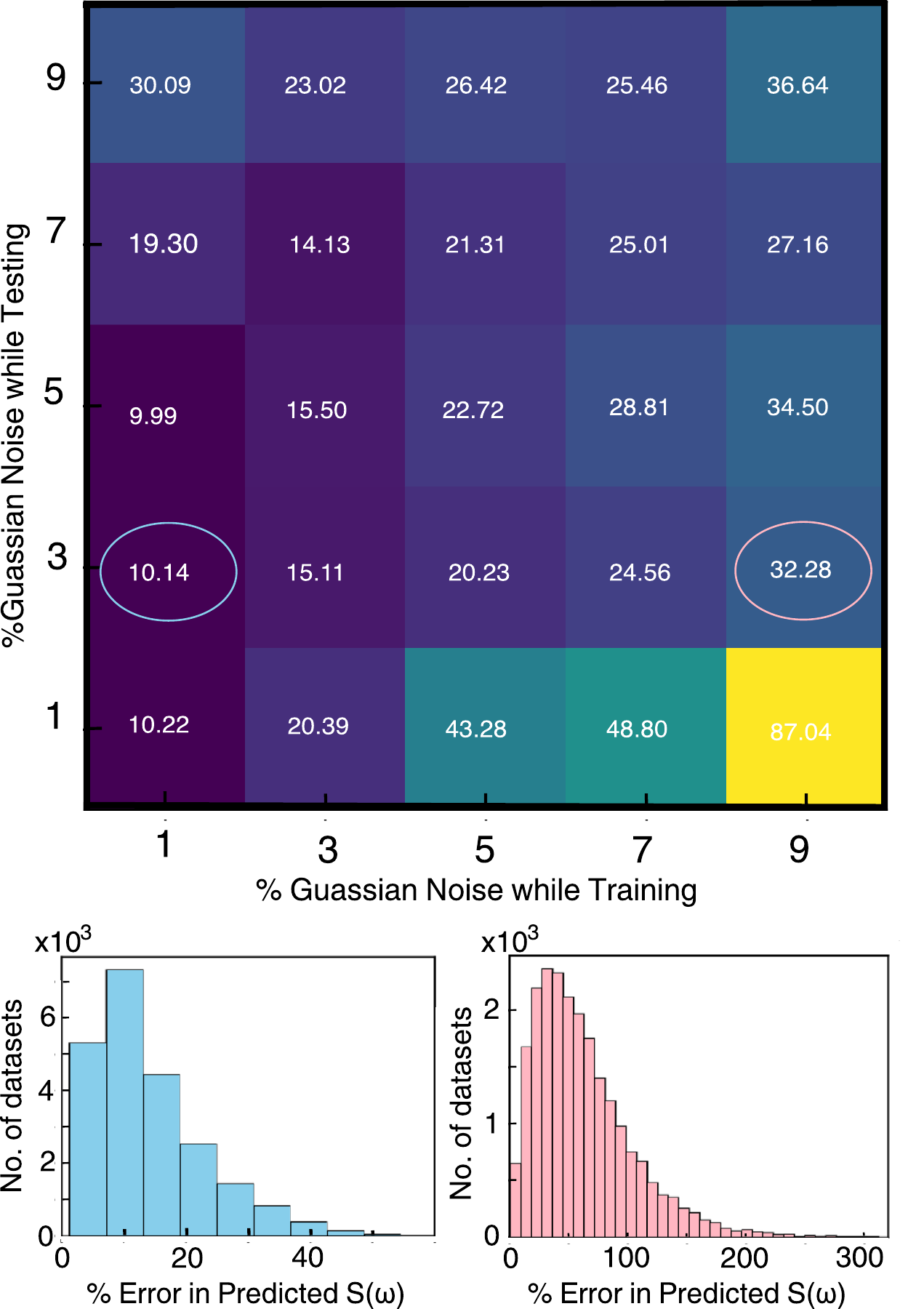}
    \caption{The heatmap displaying percentage mode values associated with the prediction errors. The $x$-axis denotes the percentage Gaussian noise used while training the model, while the $y$-axis represents the percentage Gaussian noise employed during the testing phase. The histograms at the bottom show the distribution of the percentage errors in the individual $S(\omega)$ predictions for the two scenarios specified in the heatmap.}
    \label{fig:error-analysis}
\end{figure}

The neural network-based approach to noise spectroscopy offers significant benefits over traditional methods, particularly for cases where the signal-to-noise ratio is relatively low due to a high-speed measurement protocol with a small number of shots. To demonstrate this advantage, we performed a systematic analysis of how the prediction errors scale with the unavoidable experimental noise. We trained five different networks using the same underlying training data, but with varying levels of synthetic experimental noise (i.e. Gaussian noise) ranging from 1\% to 9\%. We then used the test data having different degrees of synthetic experimental noise added to them to obtain predicted noise spectra for each of these combinations. It is important to note that in the shot-noise dominated regime, these experimental noise levels correspond to an 81-fold change in the experimental duration. Subsequently, we analyze the distribution of the errors for each case to extract the mode value; the individual mode values are depicted in the heatmap shown in Fig. \ref{fig:error-analysis}, whereas the histograms of the errors for the two specific cases highlighted in the heatmap are shown in the bottom panel. The analysis supports the following major advantages of the neural network based approach -

\begin{figure*}[!ht]
    \centering
    \includegraphics[width=7in]{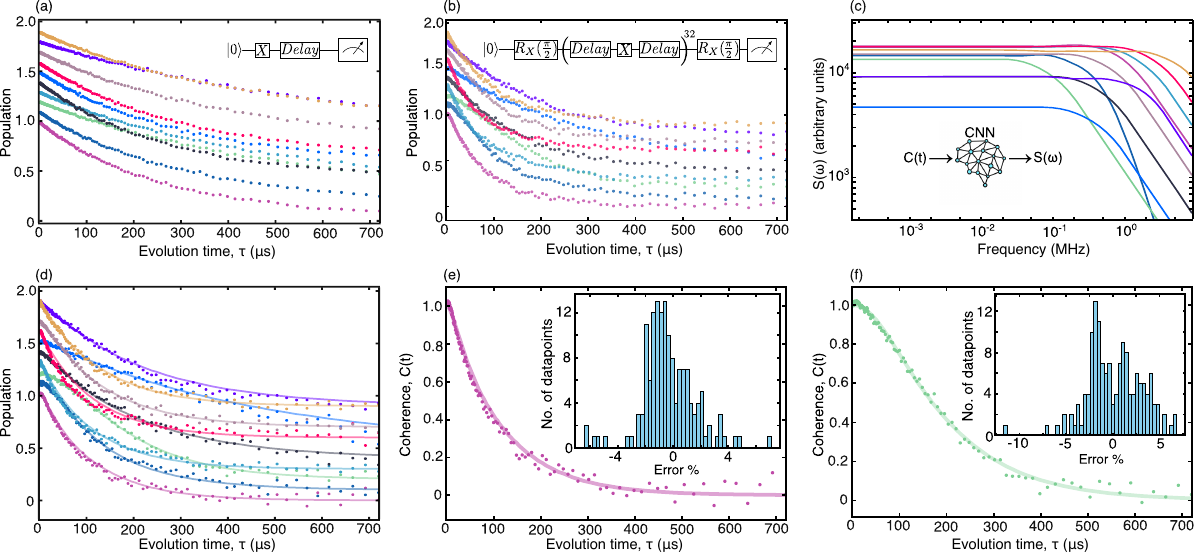}
    \caption{(a), (b),and (d) display $T_1$, $T_2$ and $T_{\phi}$ data respectively for 10 qubits on IBM Osaka. Each curve represents a qubit, with a vertical shift of 0.1 units for enhanced visibility. Dots indicate the experimental data, while in (d) the solid lines represent the network's predictions. The extracted noise spectra are shown in subplot(c). For clarity, (e), and (f) show experimental and predicted decoherence curves for 2 qubits with the absolute error histogram in the inset.}
    \label{ibm_results}
\end{figure*}

\begin{enumerate}
    \item First, the neural network trained with intermediate levels of experimental noise, e.g. $3\%$ noise, is extremely resilient to the changes in signal-to-noise ratio. This signifies that the experimenter can reduce the measurement time significantly without a substantial reduction in prediction accuracy. Moreover, this also means that the experimenter need not precisely replicate the levels of experimental noise used during training. This robustness against the noise is a characteristic feature of convolutional neural networks, which are often used for image denoising.
    \item Secondly, classical optimizers are extremely sensitive to initial guesses and frequently get trapped at local minima. On the contrary, it is known that over-parametrized neural networks tend to obtain global minima with high likelihood \cite{Choromanska2015}. Histograms with a well-defined maximum observed in the case of the neural network (Fig. \ref{fig:error-analysis}) as opposed to a distribution that appears to be reasonably random in the case of classical optimizers (see Fig. 3 in the Supplementary Material) indeed validate this claim.  
    \item Finally, the prediction speed of neural networks is undeniably superior to that of classical optimizers. Classical optimizers are known to exhibit sluggish performance and convergence issues, especially in the context of multivariate functions.
\end{enumerate}

Building on these validations, we carried out neural network–enabled quantum noise spectroscopy on a physical quantum device as discussed in the following section.

\subsection{\label{sec:level4.2}Network Performance on IBM Qubits}

We tested our noise spectroscopy methodology on IBM's superconducting qubits, specifically on the 127-qubit device, IBM Osaka. We selected specific qubits with coherence times $T_2$ between 150 $\mu s$ and 300 $\mu s$. The frequency range probed was determined by the total evolution time of the qubit --- from 2 $\mu s$ to 720 $\mu s$, which allowed sufficient coverage of both the low and the high-frequency regimes. A CPMG-32 pulse sequence was used with a $\pi$-pulse width of 48 $ns$.
%\textbf{Results and Analysis}: 
The results of $T_1$ and $T_2$ (CPMG-32) experiments shown in Fig.~\ref{ibm_results}(a) and (b) illustrate the decay of the population of state $\ket{1}$ (for $T_{1}$), and $\ket{0}$ (for $T_{2}$) as a function of the evolution time $\tau$. After processing the data as discussed in the previous section, these curves were fed into a trained model to extract the noise spectra presented in Fig.~\ref{ibm_results}(c). In Fig.~\ref{ibm_results}(d) the dots represent experimental data, while the solid lines correspond to the curves estimated using the network's predictions. The analysis revealed that the noise spectrum predominantly exhibited white noise characteristics, with a $1/\omega$-type profile emerging around 1 MHz, confirming the model's effectiveness in identifying dominant noise features. The model accurately predicted the noise spectrum for each qubit, as evidenced by a minimal prediction error in the coherence functions reconstructed from the extracted noise spectra; two example curves are shown in Fig.~\ref{ibm_results}(e) and (f). The histograms of the error distribution are plotted in the insets of Fig.~\ref{ibm_results}(e) and (f), indicating that the distribution is symmetric around zero, suggesting, yet again, that the errors were mostly limited by the experimental noise.

\subsection{\label{sec:level4.3}Optimization of Dynamical Decoupling Pulse Sequence}
\begin{figure*}[!ht]
    \centering  \includegraphics[width=0.95\linewidth]{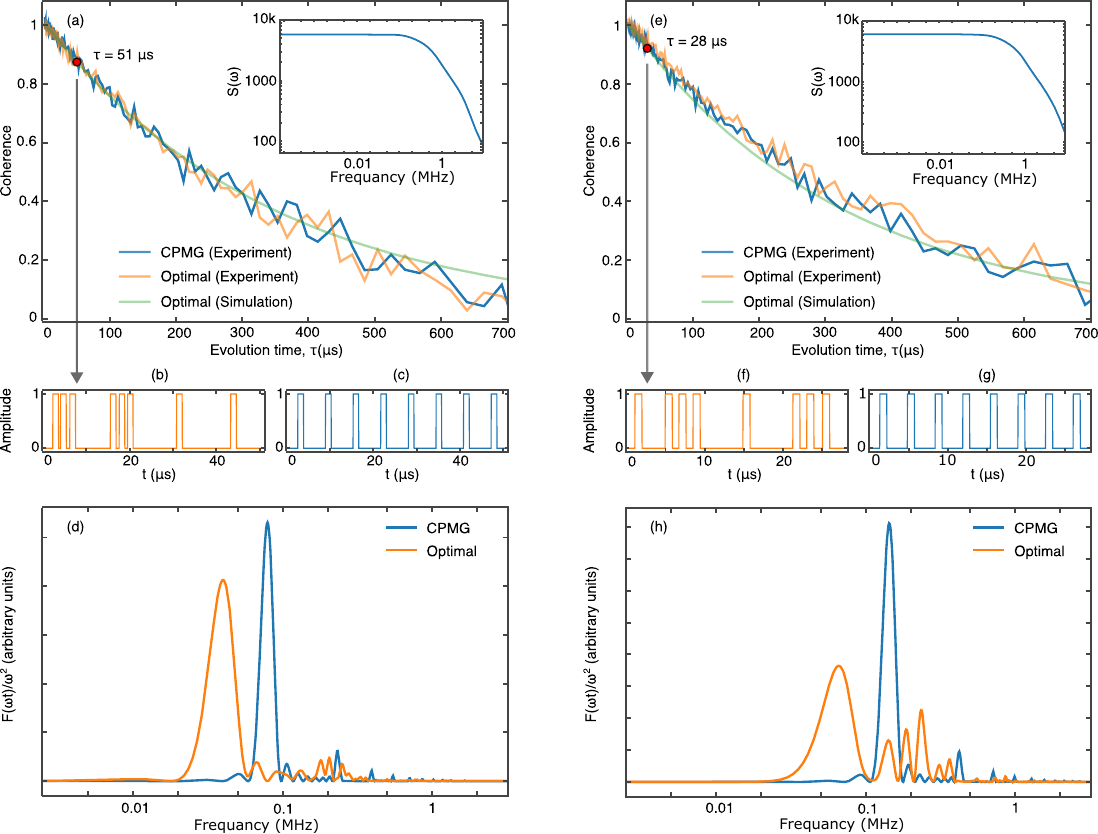}
    \caption{
    The first and second columns correspond to two different qubits from IBM-Osaka. (a), (e) Decoherence curves $C(t)$ and noise spectra $S(\omega)$ associated with the given qubit. The coherence curve is obtained by applying CPMG-8 and optimal pulse sequences ($n=8$). The blue and orange lines show the experimental coherence curves, while the green line shows the simulated optimal coherence curve. (b), (f) CPMG-8 and (c), (g) optimal pulse sequences for the given evolution time (pulse-widths are not to the scale). (d), (h) Filter functions $F(\omega t)$ corresponding the pulse sequences.}
    \vspace{0.1in}
    \label{optimalDD}
\end{figure*}

The predicted noise spectra were then used to develop the optimal sequences with varied number $n$ of $\pi$-pulses. Based on the $1/\omega$ frequency cut-off in the noise spectra, we estimated that the $n=8$ $\pi$-pulse sequences should provide the maximum advantage (see SI Section 3 and the figure therein). Our experimental results, obtained on IBM Osaka quantum processor for two different qubits are displayed in Fig.~\ref{optimalDD}(a) and (e) with the underlying noise spectra plotted in the insets. The results reveal two notable outcomes. Firstly, the customized pulse sequences shown in Fig.~\ref{optimalDD}(b) and (f) do not show noticeable improvement as compared to traditional protocols displayed in Fig.\ref{optimalDD}(c) and (g). This is due to the fact that the noise spectra are heavily white noise dominated and, consequently, the modified filter functions shown in Fig.\ref{optimalDD}(d) and (h) are unable to suppress the overlap. Secondly, despite implementing non-trivial pulse sequences, the experimental decoherence curves closely match with those predicted via the optimization algorithm. This authenticates the accuracy of the noise spectra predicted by the network. The overall methodology is, thus,  a step forward in the noise-adapted optimal quantum error mitigation techniques. The effectiveness of the optimized sequences should certainly be notable for other qubit systems where white noise is not the leading type of the environmental noise.

\subsection{\label{sec:level4.4}Rapid Time Dependent Noise Spectroscopy}

\begin{figure*}[!ht]
    \centering
    \includegraphics[width=\linewidth]{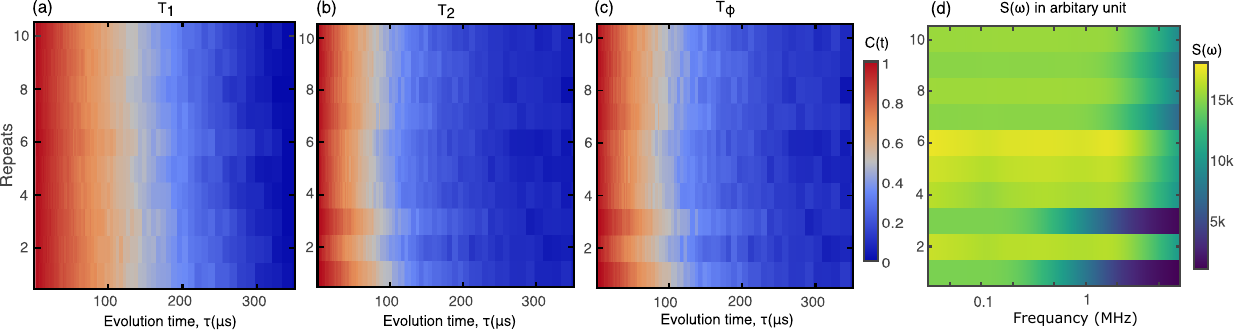}
    \caption{(a)---(c) Heatmaps displaying time-dependent $T_1$, $T_2$, and $T_\phi$ datasets for 10 repeats on the same qubit.
    (d) Heatmap of the predicted noise-spectra associated with each repeat.}
    \label{fig:heatmap}
\end{figure*}

In order to capture the time-evolution of the noise around IBM qubits, we performed time-dependent noise spectroscopy experiments on a qubit. To achieve this, ten $T_1$ and ten $T_2$ measurements were performed consecutively while keeping the experimental parameters unchanged. Note that each run consists of 4000 shots to attain sufficient signal-to-noise ratio which required approximately 10 minutes to complete. The results displayed in Fig.\ref{fig:heatmap}(a) and (b) demonstrate that we are indeed able to record the rapid changes in $T_1$ and $T_2$ curves. Moreover, the extracted noise spectra presented in Fig.\ref{fig:heatmap}(c) indicate that the discontinuous jumps in the dephasing and relaxation data are neither time-correlated nor entirely interdependent. These observations confirm that our noise spectroscopy model is a very effective tool to extract quasi-instantaneous noise profiles. 

\section{\label{sec:level5}Outlook} 

Owing to the universality of the technique discussed here, we anticipate that this study will help to examine and benchmark various quantum systems in the near future. We expect that the noise characterization methodology discussed here will provide useful insights to improve the engineering aspects of quantum systems. The fidelity of the current methodology can be improved by taking into account advanced gates \cite{Baum_RL}, or DRAG pulses \cite{Werninghaus2021, hyyppä_2025}, which are optimized to reduce leakage error, instead of square pulses. This should be possible owing to the recent advancements in the filter function formalism, which allows for analytical representation of arbitrary quantum control sequence \cite{Soare2014, Pascal_2021, Teerawat_21, Isabel_2022}. Moreover, if this methodology is employed in conjunction with appropriate instrumentation such as Field Programmable Gate Arrays (FPGAs) \cite{PhysRevX.14.041056}, then instantaneous error mitigation strategies can potentially be developed.

Finally, it is important to emphasize that the decoherence functional, $\chi(t)$, which depends on the knowledge of the environmental noise spectrum, is intricately  linked to the error rates of the various  noise operators in the Kraus picture. Since the Kraus formalism is commonly employed in  the design and implementation of QEC protocols,  having precise knowledge of the decoherence functional opens up new avenues for noise-adapted error correction protocols \cite{PhysRevResearch.6.043034}.  Indeed, QEC protocols can then be tailored to deal with a specific spectral range of noise. With the help of a trained model, time-dependent quantum error correction protocols can become a foreseeable reality.
\\
\section{Acknowledgements}
B. G. and V. J., contributed equally to this work. S.D. thanks the Indian Institute of Technology, Madras, India, for its seed and initiation grants. The authors acknowledged the use of facilities supported by a grant from the Mphasis F1 Foundation given to the Centre for Quantum Information, Communication, and Computing (CQuICC). Finally, the authors appreciated the use of IBM Quantum for this work. The views expressed were those of the authors and did not reﬂect the oﬃcial policy or position of IBM or the IBM Quantum team.

\textbf{Data Availability} \par
The relevant experimental data and the trained model parameters are available from the authors. 

\textbf{Code Availability} \par
The code for training data generation and the model construction is provided in the \href{https://github.com/bhaveshgupta1605/Noise-Spectroscopy-ML-project}{public repository}.

\bibliography{bib}

\appendix

\section{\label{sec:level6.1}Noise Model: Kraus Operator Picture}
Kraus formalism provides a comprehensive framework for modeling noise by representing the system's evolution through a set of operators, known as Kraus operators, that describe the impact of noise on the quantum system. This formalism allows for a clear and systematic analysis of various types of noise processes, including decoherence and dissipation, by translating them into a mathematical language that is both intuitive and versatile.

The evolution of a quantum system, such as a qubit, can be known mathematically via \textit{completely positive trace preserving} (CPTP) maps. If the decoherence interactions is known, we can analytically write a map which gives us the final density matrix of our system \cite{Nielsen_Chuang}.
Let $\rho_{S} \otimes \Phi_{E}$ be the initial unentangled state of the system and environment, and let $U_{SE}$ denote the unitary time-evolution operator that arises due to system---environment interaction. Then, the final state of the system alone can be obtained as follows:
\begin{equation}
  \mathcal{E}(\rho_S)=Tr_E[U_{SE}(\rho_S\otimes\Phi_E)U_{SE}^\dagger]  
\end{equation}
where the physical process map $\mathcal{E}$ on the system operator Hilbert space $\mathcal{H}_S$ must be completely positive (CP).
%: $\mathcal{E}(\rho)>0$, $\forall$ $\rho>0$ $\in \mathcal{H}_S$, where $(\mathcal{E}\otimes\mathbb{I})$ is a positive map for any possible extension $\mathcal{H}_S\otimes\mathcal{H}_E$. 
The complete positivity of the map implies the existence of an operator--sum decomposition for the map \cite{Nielsen_Chuang}, of the following form
\begin{equation}
 \mathcal{E}(\rho)=\sum_{i=1}^NE_i\rho E_i^\dagger
\end{equation}
This is known as Choi--Kraus--Sudarshan operator--sum representation, where $\{E_i\}$ are a set of operators known as Kraus operators on the state space of the system. The operators satisfy a completeness relation that arises from the requirement that the trace of $\mathcal{E}(\rho)$ be equal to one. Such maps are said to be trace preserving (TP): $Tr[\mathcal{E}(\rho)]=1 \implies \sum_i E_i^\dagger E_i = I_s$. 
Knowing the map $\mathcal{E}$ is equivalent to knowing the Kraus operators $\{E_i\}_{i=1}^N$. Here, $N$ depends on the number of basis states used to define the environment. 
%A mathematical theorem, known in the community, suggests at least $d^2$ Kraus operators is enough to know system evolution, where $d$ is the dimension of the system.\\

We will now describe two examples of such CPTP maps --- also referred to as \textit{quantum channels} \cite{Nielsen_Chuang}--- which become important in the context of superconducting qubits. The first is the so called \textit{amplitude damping} channel, which characterizes the effects due to loss of energy from a quantum system. Specifically, it describes energy dissipation in a two-level system. Let $|0\rangle$ denote the ground state and $|1\rangle$ some excited state of a qubit. Then, the amplitude damping channel denoted as $\mathcal{E}^{AD}=\{E_0^{AD}, E_1^{AD}\}$, is described by the following pair of Kraus operators \cite{Nielsen_Chuang}.
\begin{align*}
    E_0^{AD}&=\frac{1}{2}[(1+\sqrt{1-\gamma}I+(1-\sqrt{1-\gamma}\sigma_z))]\\&=\begin{pmatrix}
    1 & 0 \\ 0 & \sqrt{1-\gamma}
\end{pmatrix}, \\ E_1^{AD}&=\frac{\sqrt{\gamma}}{2}[\sigma_x+i\sigma_y]=\begin{pmatrix}
    0 & \sqrt{\gamma} \\ 0 & 0
\end{pmatrix}
\end{align*}
Here, $\{I,\sigma_x,\sigma_y,\sigma_z\}$ is the Pauli basis and $\gamma$ is the probability of a transition from the excited state to the ground state. As a function of time, this transition probability can be expressed as $\gamma = 1-e^{-\Gamma_1t}$, where $\Gamma_1 = 1/T_1$ is the inverse of the relaxation time $T_1$.\\

Another example of a noise process is that of \textit{phase damping} channel, which describes the loss of relative phase information between the energy eigenstates. Then phase damping channel denoted as $\mathcal{E}^{PD}=\{E_0^{PD}, E_1^{PD}\}$, is described by the following pair of Kraus operators \cite{Nielsen_Chuang} given by 
\begin{align*}
     E_0^{PD}&=\sqrt{p}I=\begin{pmatrix}
    \sqrt{p} & 0 \\ 0 & \sqrt{p}
\end{pmatrix}, \\  E_1^{PD}&=\sqrt{1-p}\sigma_z=\begin{pmatrix}
    \sqrt{1-p} & 0 \\ 0 & -\sqrt{1-p}
\end{pmatrix}  
\end{align*}
Note that phase damping is also often referred to as the \textit{phase flip} channel \cite{Nielsen_Chuang}. Here, $p$ is the probability of not a phase flip. As a function of time, this probability can be expressed as, $p=\frac{1}{2}(1+e^{-\Gamma_\phi t})$, where $\Gamma_\phi = 1/T_\phi$ is the inverse of the relaxation time $T_{\phi}$.\\

Consider an arbitrary single-qubit state, written in the standard basis as, $|\psi\rangle = \alpha |0\rangle + \beta|1\rangle$. Let $\rho = |\psi\rangle\langle\psi|$ denote the density operator corresponding to this state. Then, the density operator after the combined action of both these noise processes is given by 
\begin{align}\label{krauss_density_matrix}
\rho &= \begin{pmatrix}
        |\alpha|^2 & \alpha \beta^* \\ \alpha^*\beta & |\beta|^2    \end{pmatrix}\\
        % \xrightarrow{noise}
        \rho_{noisy}&=\mathcal{E}^{AD}\circ\mathcal{E}^{PD}(\rho)\\
        &=\mathcal{E}^{PD}\circ\mathcal{E}^{AD}(\rho)\\
        % \nonumber \\ & 
        \rho_{noisy}&=\begin{pmatrix}
    1+(|\alpha|^2-1)e^{-\Gamma_1t} & \alpha\beta^*e^{-\frac{\Gamma _1}{2}t} e^{-\Gamma _{\phi}t} \\ \alpha^*\beta e^{-\frac{\Gamma _1}{2}t} e^{-\Gamma _{\phi}t} & |\beta|^2e^{-\Gamma_1t}
\end{pmatrix}
\end{align}
It does not matter in which order the noise acts since both noise processes are independent. In superconducting qubits, the dominant noises are amplitude damping, also referred to as energy relaxation or longitudinal relaxation, and phase damping, also referred to as pure dephasing. 

\section{\label{sec:level6.2}Neural Network} 
% \begin{widetext}
%     \begin{center}
\begin{table*}[h]
\centering
\begin{tabular}{|c|c|c|c|c|}
    \hline
    \textbf{Layer (Type)} & \textbf{Output Shape} & \textbf{Parameters} & \textbf{Activation} & \textbf{Notes} \\ 
    \hline
    Input Layer & (xtrain size, 1) & 0 & - & Accepts one-dimensional data \\ 
    \hline
    Conv2D (1) & (None, xtrain size, 40) & 1040 & ReLU & 40 filters, 5x5 kernel, same padding \\ 
    \hline
    MaxPooling2D (1) & (None, xtrain size, 40) & 0 & - & Pool size = `pool size', same padding \\ 
    \hline
    Conv2D (2) & (None, xtrain size, 40) & 40040 & ReLU & 40 filters, 5x5 kernel, same padding \\ 
    \hline
    MaxPooling2D (2) & (None, xtrain size, 40) & 0 & - & Pool size = `pool size', same padding \\ 
    \hline
    Conv2D (3) & (None, xtrain size, 40) & 40040 & ReLU & 40 filters, 5x5 kernel, same padding \\ 
    \hline
    MaxPooling2D (3) & (None, xtrain size, 40) & 0 & - & Pool size = `pool size', same padding \\ 
    \hline
    Conv2D (4) & (None, xtrain size, 40) & 40040 & ReLU & 40 filters, 5x5 kernel, same padding \\ 
    \hline
    MaxPooling2D (4) & (None, xtrain size, 40) & 0 & - & Pool size = `pool size', same padding \\ 
    \hline
    Conv2D (5) & (None, xtrain size, 80) & 80080 & ReLU & 80 filters, 5x5 kernel \\ 
    \hline
    UpSampling2D (1) & (None, xtrain size, 80) & 0 & - & Upsampling size = `pool size' \\ 
    \hline
    Conv2D (6) & (None, xtrain size, 160) & 160160 & ReLU & 160 filters, 5x5 kernel \\ 
    \hline
    UpSampling2D (2) & (None, xtrain size, 160) & 0 & - & Upsampling size = `pool size' \\ 
    \hline
    Conv2D (7) & (None, xtrain size, 320) & 320320 & ReLU & 320 filters, 5x5 kernel \\ 
    \hline
    Conv2D (8) & (None, xtrain size, 1) & 8001 & ReLU & 1 filter, 5x5 kernel \\ 
    \hline
    Flatten & (None, xtrain size) & 0 & - & Converts to 1D vector \\ 
    \hline
    Dropout & (None, xtrain size) & 0 & - & Dropout rate = `dropout rate' \\ 
    \hline
    Dense (1) & (None, 501) & xtrain size * 501 + 501 & Linear & Final output layer \\ 
    \hline
    \multicolumn{5}{|c|}{Total Parameters: 769,721} \\ 
    \hline
\end{tabular}      
\caption{Summary of the Network. Here typical xtrain size = 150, `pool size' = 2, and `dropout rate' = 0.05.} 
\label{table_2}
\end{table*}
%     \end{center}
% \end{widetext}
We used TensorFlow Keras Python module to train our convolutional neural network and to perform the noise spectra predictions from the experimentally obtained decoherence curves. The autoencoder-type network, as described in Table \ref{table_2} alternates between convolutional and pooling operations to progressively extract features from the input data, and the upsampling restores the data's original dimensions while enhancing feature representation. The final dense layer aims to produce a linear output appropriate for regression or other continuous output tasks. Our network is trained for approximately 75 epochs, achieving an accuracy loss lower than 4\%. The training and validation losses, measured as mean absolute error, are monitored over epochs. 

\begin{figure}[h]
    \centering
    \includegraphics[width=\linewidth]{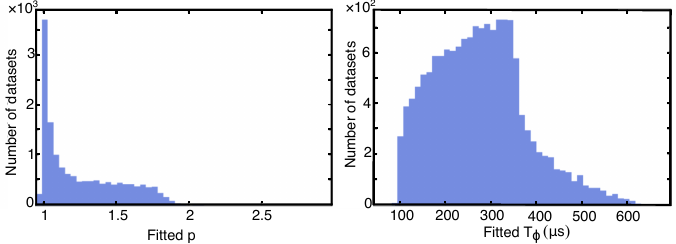}
    \caption{Distribution of fitted $p$ and $T_\phi$ in the training dataset}
    \label{training_data}
\end{figure}
Furthermore, to obtain a high degree of prediction accuracy, it is essential to constrain the training data appropriately. We fit the generated decoherence curves using a stretched exponential function:
\begin{equation}\label{streched-exponential}
\tilde{C}(t) = e^{-(t/T_\phi)^p}
\end{equation}
Although, the simulated curves cannot be exactly fitted with Eq.\ref{streched-exponential}, this method provides a good measure to filter the training data. SI Figure \ref{training_data} shows the final distribution of the stretching factor $p$, and of inverse decay rate $T_\phi$ in the training dataset. All the experimental curves lie well within the chosen bounds.

\section{Dynamical Decoupling Optimization}
\begin{figure}[h]
    \centering
    \includegraphics[width=\linewidth]{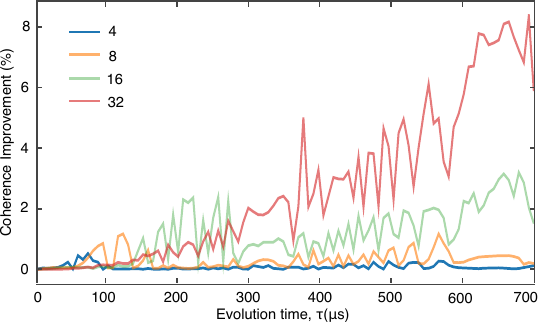}
    \caption{$\%$ Coherence Improvement for a randomly chosen qubit as a function of evolution time for CPMG pulse sequences with $n$=4,8,16, and 32 $\pi$-pulses each having a width of 48 ns.}
    \label{CPMG-8}
\end{figure}

To determine the optimal number of pulses for practical implementation, we performed a systematic analysis comparing CPMG sequences with a varied number of pulses $(n = 4, 8, 16, 32)$. Fig.\ref{CPMG-8} demonstrates the $\%$ coherence improvement as a function of evolution time for these different sequences. Supposing that the gain is relevant only when it occurs at short evolution times ($<T_2$), it can be postulated that 8 $\pi$-pulse sequences are close to optimal. The $\%$ improvements observed at longer evolution times, specifically for $n=16,32$ are practically ineffective.

\section{\label{sec:level6.3} Classical Optimization vs Machine Learning (ML) based Approach}

In order to validate the superiority of our approach we compared the results shown in Figure 2 of the manuscript with the state-of-the-art classical optimizers as depicted in the SI Figure 3. For this purpose, we implemented brute force optimization using the functional form of $S(\omega)$ described in Section 3 of the manuscript. We used the Constrained Optimization BY Linear Approximation (COBYLA) algorithm provided by SciPy. We analyzed 200 test datasets having the same percentage of experimental noise and repeated the procedure for various noise percentages. For the ML-based approach, the error distributions (presented in Figure 2 of the manuscript) have a well-defined maxima; therefore, we extracted the mode value associated with each histogram and considered the standard deviation around the value to generate the error bars. Whereas, the error distribution in the case of classical optimization does not show a clear trend, and thus, we use the mean value and standard deviation around the mean to analyze the data. The comparative plot evidently shows that the ML-based technique outperforms the classical optimization approach even in the case of the data with a low signal-to-noise ratio.
\begin{figure}[ht]
    \centering
    \includegraphics[width=\linewidth]{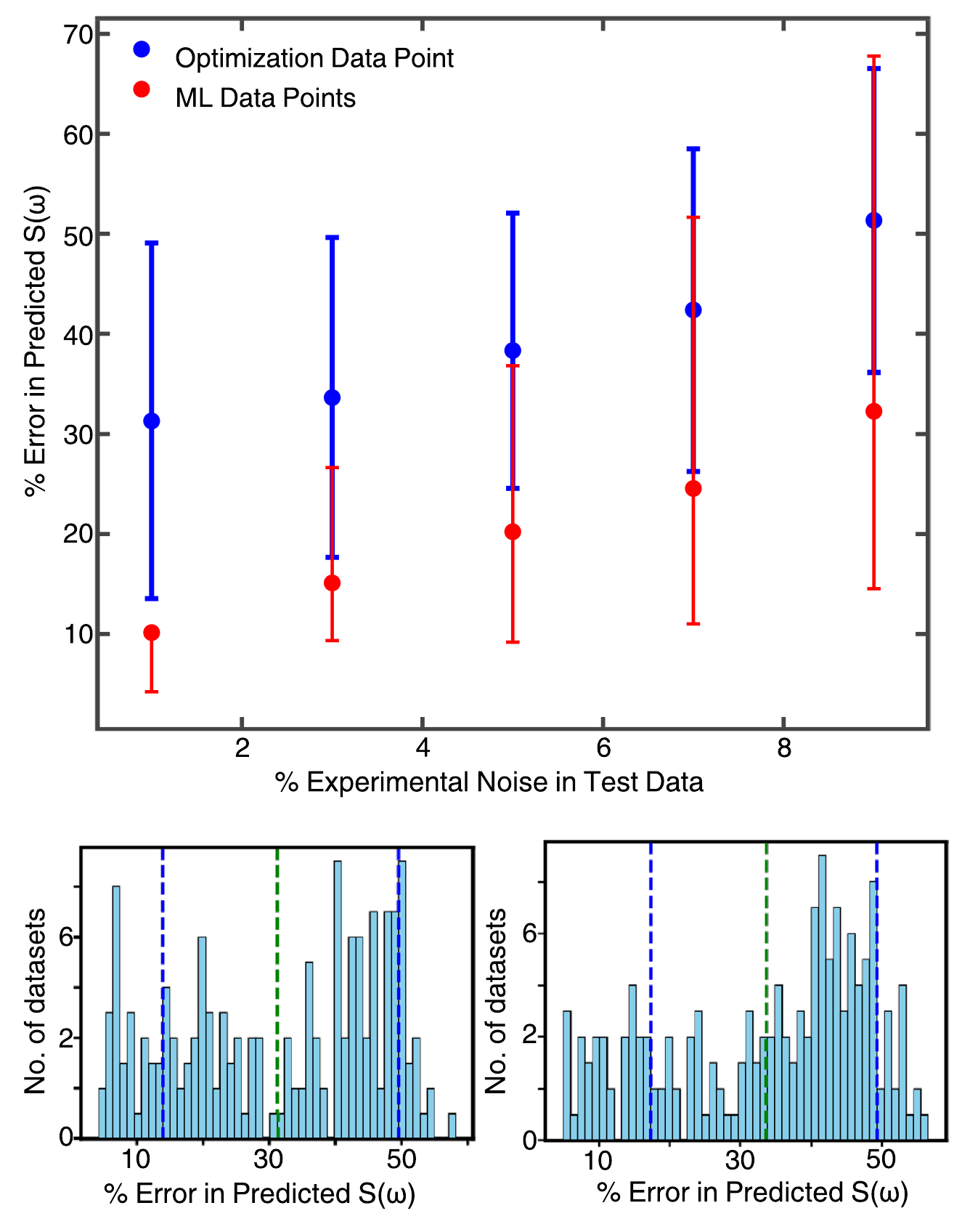}
    \caption{$\%$ error in the predicted $S(\omega)$ for the two distinct methodologies. The histograms at the bottom, show the distribution of the $\%$ errors in the individual predictions for the classical optimization protocol. These histograms are shown for the datasets comprising the lowest and highest experimental noise. The green dashed line represents the mean, while the blue dashed line represents the $\%$ error 1 standard deviation away from the mean value.}
    \label{fig:CO-ML compare}
\end{figure}

\end{document}